# Optimizing type-I polarization-entangled photons

Radhika Rangarajan,[1*] Michael Goggin,[2] and Paul Kwiat[1]

[1]*Department of Physics, University of Illinois, 1110 Green Street, Urbana, IL 61801, USA*
[2]*Department of Physics, Truman State University, Kirksville, MO 63501*
[*]*rangaraj@illinois.edu*

**Abstract:** Optical quantum information processing needs ultra-bright sources of entangled photons, especially from synchronizable femtosecond lasers and low-cost cw-diode lasers. Decoherence due to timing information and spatial mode-dependent phase has traditionally limited the brightness of such sources. We report on a variety of methods to optimize type-I polarization-entangled sources — the combined use of different compensation techniques to engineer high-fidelity pulsed and cw-diode laser-pumped sources, as well as the first production of polarization-entanglement directly from the highly nonlinear biaxial crystal $BiB_3O_6$ (BiBO). Using spatial compensation, we show more than a 400-fold improvement in the phase flatness, which otherwise limits efficient collection of entangled photons from BiBO, and report the highest fidelity to date (99%) of any ultrafast polarization-entanglement source. Our numerical code, available on our website, can design optimal compensation crystals and simulate entanglement from a variety of type-I phasematched nonlinear crystals.

---

**1. Introduction: Brightness limitations of type-I entanglement sources**

Scalable optical quantum computation and quantum communication require ultra-bright sources of optical qubits, especially entangled photons. High-quality polarization-entangled states have historically been produced via the nonlinear process of spontaneous parametric downconversion (SPDC) [1], in which one high-energy pump photon splits into two lower-energy daughter photons (called the signal and idler); recently sources based on four-wave mixing have been developed [2-3]. The brightness of SPDC-based entanglement sources is limited in practice by decoherence, and fundamentally by the nonlinear dielectric tensor of the SPDC crystal used. Here, we present ways to design ultra-bright type-I [4] sources of entangled photons by combining multiple decoherence-compensation techniques, as well as by incorporating an unconventional highly nonlinear biaxial SPDC crystal, bismuth triborate ($BiB_3O_6$, BiBO).

The dominant decoherence mechanisms in polarization-entangled sources depend on the source specifications, which in turn depend on the particular application. For instance, ultrafast pulsed entanglement sources provide critical timing information and enable synchronization, making them essential for various quantum information processing protocols, including quantum teleportation [5-6] and optical quantum computing [7-8]. Additionally, several recent schemes for engineering entangled-photon modes exploit the range and control of pump bandwidth facilitated by a pulsed femtosecond laser [9-11]. In contrast, given their simplicity, low costs and portability, cw-diode lasers appear attractive as pumps for applications such as quantum cryptography and investigating fundamental physics in undergraduate laboratories [12-13]. However, decoherence in entanglement sources employing ultrafast or diode-laser pumps causes a tradeoff between source brightness and fidelity, making it challenging to create an efficient high-fidelity entanglement source using these pumps. For example, sources based on type-II phasematching, thus far the predominant method for ultrafast entanglement generation, are nevertheless limited by fundamentally small solid angles over which entanglement persists, or require interferometric configurations [14-19]. In contrast, type-I entanglement sources are advantageous because of their comparatively high brightness, stability and ease-of-alignment [20]. Such sources have nevertheless traditionally been limited by reduced entanglement with larger collection irises and increasing pump bandwidths. Here we consider the two main dephasing mechanisms that degrade type-I polarization entanglement: emission-angle dependent phase, and pump-frequency dependent phase, characteristic of low coherence-time pumps such as ultrafast and free-running diode lasers. The entanglement quality is typically recovered by strong filtering in the extra degree of freedom (e.g., narrow irises and spectral filters) thereby drastically reducing the collection efficiency [21-23]. However, using a spatial-phase compensation technique, one can drastically increase the brightness for type-I sources without sacrificing source fidelity [24]. Specifically, by successfully correcting for the directional dependence of the relative phase between the different polarization components, the coincidence collection efficiency was increased by more than 50 times while maintaining a fidelity >97% for entangled sources pumped with monochromatic cw lasers. Here, we extend this technique to both ultrafast-pumped and cw-diode laser-pumped entanglement sources, achieving brightness enhancements up to 400.

In addition to the spatial decoherence just mentioned, downconversion sources pumped with ultrashort pulses are further plagued with decoherence arising from different pump-frequency components. One can use temporal precompensation techniques to mitigate this phenomenon [23]. In fact, they can also improve the fidelity of polarization-entangled photons generated using cw-diode lasers that contain a range of pump frequencies, due to short coherence times, mode-hopping, etc. [12, 25-26]. Here, we present optimized solutions to improve the fidelity and brightness from a variety of type-I polarization-entangled sources — ultrashort and cw-diode pumped, for both degenerate and, for the first time, non-degenerate entanglement — achieved by combining temporal and spatial compensation techniques. We show that these techniques can be applied simultaneously, as long as one correctly accounts for the total effects of both the compensators. We have thus realized the highest entangled state fidelity (99%) [27] for a cw-diode laser-pumped degenerate source, as well as the highest reported entanglement concurrence [28] (98%, previously limited to 94% [23] for type-I sources) for a polarization-entanglement source pumped by an ultrashort pulsed laser.

Aside from these correctable decoherence effects, the downconversion efficiency in any experiment is *fundamentally* limited by the nonlinear dielectric properties of the SPDC crystal itself. For example, the uniaxial crystal barium borate ($\beta$-BaB$_2$O$_4$ or BBO with effective nonlinear coefficient $d_{eff}$ ~1.75 pm/V) has been almost ubiquitously employed for polarization-entanglement generation, although quasi-phasematching in materials such as periodically-poled potassium titanyl phosphate (KTiOPO$_4$, KTP $d_{eff}$ ~ 3 pm/V) are becoming more popular [29]. Here we present the first direct entanglement results from the promising newly developed nonlinear optical crystal BiBO, which has exceptionally high nonlinearity (>3 pm/V), UV transparency, high damage threshold (comparable to BBO) and inertness to moisture (nonhygroscopic) [30-32]; moreover, as a biaxial crystal, BiBO offers versatile phasematching characteristics and broadband angle-tuning at room temperature [31]. Such a combination of properties makes this crystal highly attractive for frequency conversion in the UV, visible and IR; for example, BiBO is superior to BBO for second-harmonic-generation [33]. Nevertheless, while biaxial crystals such as BiBO may appear to be a better source material, their lack of the rotational symmetry present in traditionally used uniaxial crystals can significantly complicate the compensation techniques presented here, because, e.g., phasematching for a biaxial crystal depends on the azimuthal pump angle in addition to the polar pump angle. Here we show that, in spite of their increased complexity, such biaxial crystals can be used for a brighter source of polarization entanglement. Specifically, we numerically model and experimentally realize optimally compensated sources of high-quality (fidelity >99%) polarization-entangled photons using a pair of biaxial BiBO crystals. Further, our simulation code, available on our website [34], can model type-I entanglement sources from a variety of nonlinear crystals and phasematching parameters.

The remainder of the paper is arranged as follows. In Section 2 we present experimental data and numerical simulations, discussing first spatial then spectral-temporal and finally combined compensation results, for both a cw-diode laser-pumped biaxial BiBO and an ultrafast-pumped BBO polarization-entanglement source. Theoretical details, including the principles of our numerical simulations, and specific calculations for designing the compensation crystals, are further discussed in Section 3.

## 2. The two-crystal source of type-I polarization-entangled photons

We use the two-crystal geometry [20] to produce polarization-entangled photons from type-I phasematched SPDC: two adjacent thin nonlinear crystals are oriented orthogonally, such that a vertically V (horizontally H) polarized pump photon can downconvert into a pair of horizontally (vertically) polarized photons in the first (second) crystal (Fig. 1a). The downconversion processes in each crystal are coherent with one another, so pumping with photons polarized at 45° ideally generates a maximally entangled state:

$$|\psi\rangle = \frac{1}{\sqrt{2}}\left(|H_1 H_2\rangle + e^{i\phi(\omega_p, \omega_s, \omega_i, \mathbf{k}_p, \mathbf{k}_s, \mathbf{k}_i)}|V_1 V_2\rangle\right). \quad (1)$$

The relative phase φ in (1) is determined by phasematching constraints and depends on various parameters such as crystal type and length, and pump (downconversion) frequency $\omega_p$ ($\omega_s$, $\omega_i$) and momentum vector $\mathbf{k}_p$ ($\mathbf{k}_s$, $\mathbf{k}_i$). Explicit calculations for φ are presented in Section 3. The coherence between the two downconversion amplitudes generated in the adjacent nonlinear crystals can be destroyed — decreasing the amount of polarization entanglement — by correlations between the polarization and other degrees of freedom; these then effectively yield distinguishing *which-crystal* information. In particular, due to birefringence and dispersion, downconversion photons emitted at varying angles and frequencies can acquire different relative phases. Collecting multiple such states through large-diameter irises and large-bandwidth spectral filters results in averaging over the phases, leading to effective *spatial* and *spectral-temporal* decoherence, respectively. Spatial decoherence can be eliminated by directing the downconversion photons through suitable birefringent compensating crystals that have the opposite phase characteristics as that of the downconversion crystals, a technique called spatial compensation [24]. Similarly, spectral-temporal decoherence can be countered by "precompensating" the pump by passing it through a birefringent crystal before the downconversion crystals [23].

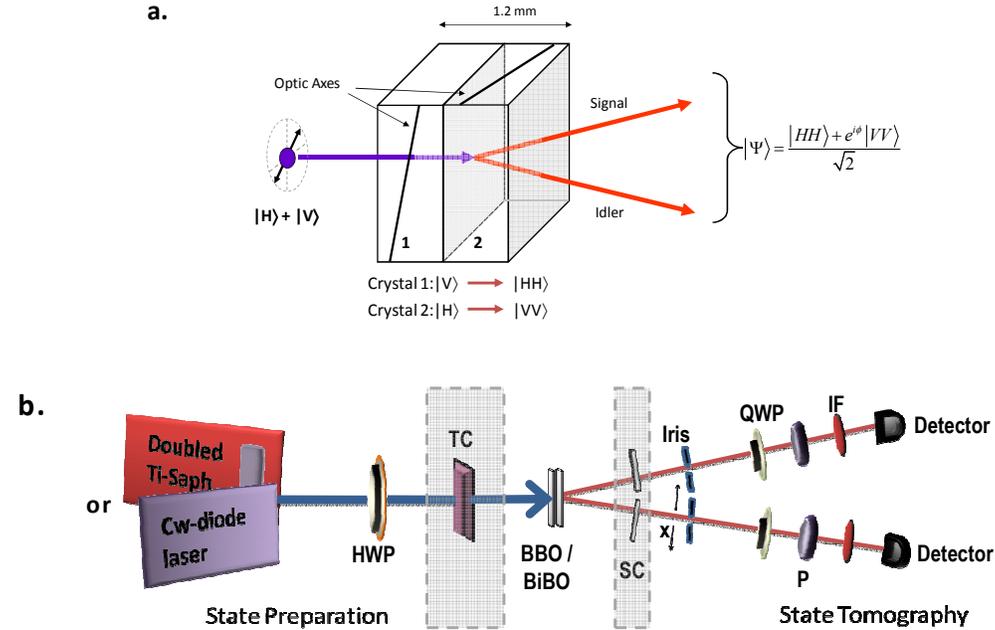

Fig. 1. **a**. Schematic showing the optic axes orientations and the corresponding downconversion photons from two orthogonally oriented type-I phasematched nonlinear crystals used to produce polarization-entangled photons. **b**. Experimental setup to generate and analyze entangled photons. 405-nm light from a doubled ultra-short Ti-Saph laser or a cw-diode laser pumps either two BBO or BiBO crystals. A half wave plate (HWP) prepares the pump state. Quarter waveplates (QWP), and polarizers (P) are used to analyze the downconverted state. 10-nm interference filters (IF) reduce background to the single-photon detectors. Birefringent spatial compensators (SC) compensate for angle-dependent phase variation. Quartz or BBO temporal compensators (TC) precompensate the pump for temporal walkoff.

We first present results showing spatial and spectral-temporal decoherence independently, and then present the combined completely compensated systems. Experimentally, we use two distinct setups (shown in Fig. 1b): in the first, two biaxial BiBO crystals are pumped by a 405-nm cw-diode laser; in the second, two BBO crystals are pumped by a 90-fs 810-nm Ti-Saph laser, frequency-doubled to obtain 405 nm. Both setups produce degenerate downconversion (405 nm $\rightarrow$ 810 nm + 810 nm); the BiBO crystals were also used to study nondegenerate downconversion (405 nm $\rightarrow$ 851 nm + 771 nm). In all cases, interference filters (10-nm FWHM) before the detectors were used to reduce background.

*2.1. Spatial decoherence and compensation results with a biaxial crystal*

As discussed above, emission-angle dependent spatial decoherence can be countered by inserting two birefringent compensators, one in each downconversion arm [35]. Given the spatial-phase characteristics of the downconversion photons generated by a particular source, the optimal length of spatial compensation crystals can be calculated for any optic-axis cut.

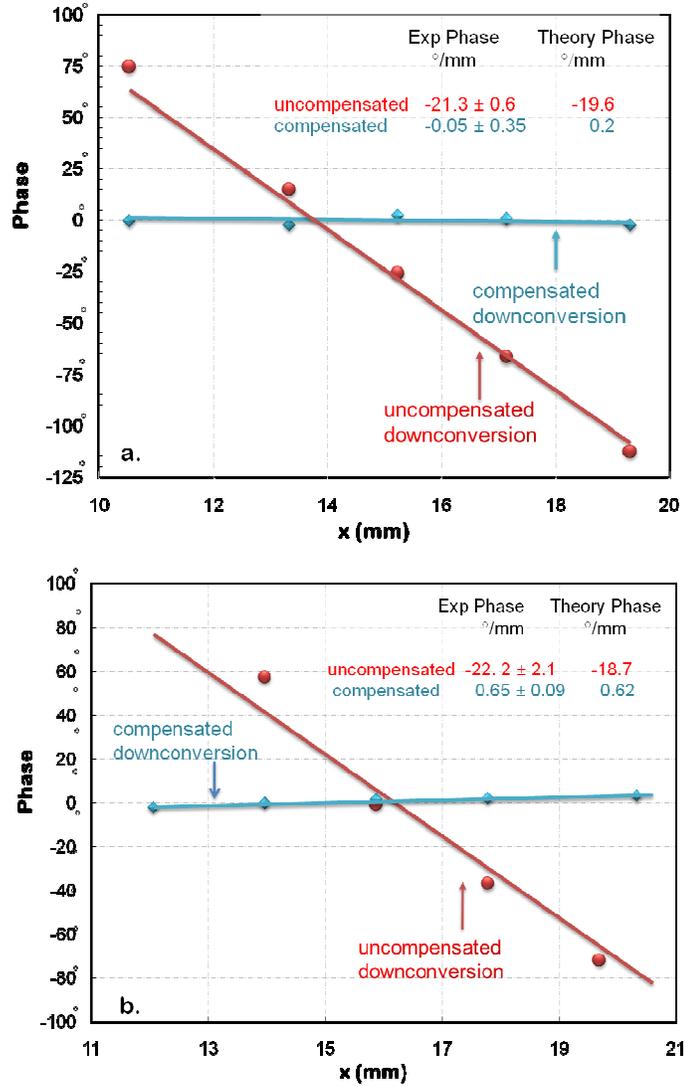

Fig. 2: Experimentally measured phase gradients for uncompensated and compensated two-crystal geometry for **a**. degenerate downconversion (405 nm $\rightarrow$ 810 nm + 810 nm) and **b**. nondegenerate downconversion (405 nm $\rightarrow$ 851 nm + 771 nm) in BiBO pumped by a cw-diode laser. The solid lines are the phase gradients predicted by our numerical simulation. The error ($\pm 0.05°$) is smaller than the data markers.

Such spatial compensation has been previously demonstrated only in uniaxial crystals (BBO) pumped with a monochromatic cw laser [24]. Here we extend this technique for the first time to spatially compensate an entangled source using biaxial crystals (BiBO), whose complex birefringent structure substantially complicates the necessary calculations of $\phi$. Our two 0.6-mm thick BiBO crystals are pumped by a 405-nm cw-diode laser. The crystals are cut at $\theta=151.7°$ and $\varphi=90°$ to produce degenerate downconversion (at 810 nm) into a cone with ~3° external opening half-angle. As shown in Fig. 2a, the uncompensated source has a large phase gradient: $\phi$ varies by more than 20° per mm in the spatial dimension x transverse to the emission direction (see Fig. 1b), measured at the collection irises (~84 cm from the downconversion crystals). The plots in Fig. 2 were obtained by translating narrow collection irises (which scans the spatial dimension x) and at each point performing a full quantum state tomography to reconstruct the two-photon state. The phase was then calculated from the resulting density matrix. Thus, collecting the photons with moderate-size irises (e.g., 5-mm diameter) would greatly decohere the polarization entanglement. To counter this, we insert a birefringent spatial compensator (245-μm thick BBO crystal cut at 33.9°) into each downconversion arm. Our calculations predict a nearly ideal phasemap, and indeed our measured phasemap is essentially flat: the spatially compensated source has a residual phase below 0.05° over 1 mm, representing a 400-fold improvement for degenerate downconversion generated using biaxial BiBO. Fig. 2b shows similar results for nondegenerate downconversion from 405 nm to 851 nm and 772 nm. Here we observe a 36-fold improvement over the uncompensated phasemap, limited by the suboptimal length of our available spatial compensators (245 μm, compared to the optimal lengths, for e.g., 280 μm for the signal and 210 μm for the idler); in any event, the degree of compensation achieved precisely matches our theoretical prediction. Note that this is also the first nondegenerate source to be spatially compensated; the flexibility of the nondegenerate operation greatly increases the utility of SPDC sources, e.g., for coupling to atomic transitions. For our second system, using ultrafast (~90 fs) pumped 0.6-mm BBO cut at 29.3°, calculations indicate that the downconversion phase slope is ~17°/mm, which is nearly perfectly compensated using two 245-μm BBO crystals cut at 33.9° as spatial compensators.

*2.2. Spectral-temporal decoherence for ultrafast / cw-diode laser pumps*

Spectral decoherence arises due to frequency-dependence of the relative phase $\phi$. It can be intuitively understood (especially in the case of a pulsed pump) in the temporal domain (hence the term *spectral-temporal* decoherence): different propagation speeds of the pump and downconversion photons within the crystals leads to temporal which-crystal information. The $|HH\rangle$ photons emitted in the first crystal are delayed by $\Delta t$ compared to the $|VV\rangle$ photons generated in the second crystal. For nondegenerate downconversion, there may also be relative delays between photons generated in the *same* crystal. If any of these relative delays is comparable to or greater than the pump coherence time, entanglement quality will be reduced (for delays between photons from the same crystal, the downconversion coherence time also affects the which-crystal information). Such spectral-temporal decoherence is extremely significant with an ultrafast pump. For example, dispersion over the ~4-nm pump bandwidth at 405 nm results in complete elimination of any polarization-entanglement when using 10-nm FWHM spectral filters. Specifically, group velocity dispersion in the two 0.6-mm BBO crystals delays the $|HH\rangle$ downconversion photons generated in the first crystal by ~253 fs relative to the $|VV\rangle$ photons from the second crystal, much more than the ~90-fs pulse duration. We used 1.9 mm of BBO cut at 29.4° to optimally compensate for the 253-fs walkoff, recovering a high quality polarization-entangled state (98.9% fidelity). In fact, for our configuration (90-fs pump + 0.6-mm crystals), higher order effects such as distinguishing dispersive broadening were calculated to be negligible; we were thus able to observe similar compensation results using 6.8-mm quartz precompensators instead. However, higher-order broadening effects would become significant for shorter pulse widths and longer crystals. For instance, we calculate that if the pump bandwidth exceeds ~10 nm, first-order correction — only compensating for group velocity delays but not pulse spreading — will yield only a 90% fidelity. Thus, while any birefringent element of appropriate thickness could be used to precompensate for the temporal walkoff to first order, using the same material as in the downconversion source itself can better compensate for higher-order effects that can be significant for increased pump bandwidths.

Spectral-temporal effects are less critical for a cw-diode laser-pumped source, but optimal precompensators can still significantly improve the measured polarization entanglement. In our system a ~0.5-nm bandwidth cw-diode laser pumps two 0.6-mm BiBO crystals phasematched for downconversion from 405 nm to 810 nm. The resulting ~600-fs delay between the $|HH\rangle$ downconversion photons born in the first crystal and the $|VV\rangle$ photons from the second crystal leads to an uncompensated tangle of only ~70%. Given the relatively small pump bandwidth of the diode laser, ~16 mm of quartz (cut with its optic axis perpendicular to the direction of propagation) can ideally compensate for this delay. Fig. 3 shows the effect of varying the delay between the $|H\rangle$ and $|V\rangle$ pump components, achieved by varying the quartz precompensator thickness, on the tangle of the resulting two-photon polarization-entangled state.

### 2.3. Joint spatial and spectral-temporal compensation

Although both the decoherence effects – spatial and spectral-temporal – occur independently, the compensation for these *cannot* in general be independent. This is because the compensation crystals themselves can cause similar effects (see Section 3), e.g., the spatial compensators can cause their own temporal walkoff of the photons. For example, the spatial compensators we used for the cw-diode laser-pumped BiBO source increase the net temporal walkoff from ~600 fs to ~640 fs, which can be compensated for with ~17.2 mm of quartz. Since it was readily available to demonstrate the (nearly) optimized spatially and temporally compensated BiBO source, we instead used two pieces of quartz tilted to have an effective thickness of ~17.9 mm as the temporal compensator. Quantum state tomography [36] was used to determine the reduced density matrix (i.e., only the polarization part) of the entangled two-photon state, for various iris sizes. From this we determined the associated two-qubit properties, such as fidelity with a maximally entangled state, and tangle. If the relative-phase variation with emission angle has truly been cancelled out, these metrics should be independent of the size of the collection irises, allowing for a quadratic increase in the brightness of the collected photon pairs with increasing iris diameters. Figure 4 shows the measured and predicted tangle as a function of iris size for both the uncompensated and spatially compensated cases, for degenerate and nondegenerate downconversion in temporally compensated BiBO. In both cases, the compensated source shows much less reduction in tangle with iris size, as predicted. The maximum tangle is less than 1, matching the prediction, due to the suboptimal length of the available temporal precompensator. As shown in Fig. 4, this effect is worse for nondegenerate pairs, because then both the spectral-temporal as well as the spatial compensators are non-ideal.

Downconversion from an ultrafast pump can be similarly compensated for spatial and spectral-temporal decoherence simultaneously. Accounting for the temporal walkoff in the spatial compensators, we used ~2.1 mm of BBO cut at 29.4° to precompensate our optimized ultrafast-pumped BBO source. By incorporating both spatial and temporal compensators, we obtained a tangle of 98% and fidelity of 99%, the highest reported thus far for polarization entanglement generated using an ultrafast pump. Figure 5 shows a plot of the density matrices obtained for all our jointly spatially and temporally compensated sources — ultrafast degenerately pumped BBO (Fig. 5a), and cw-diode laser-pumped degenerate (Fig. 5b) and non-degenerate (Fig 5c) entanglement from BiBO. Table I directly compares the measured downconversion brightness between BiBO and BBO, when they are both pumped by the cw-diode laser. BiBO's higher nonlinearity results in a 3-fold brightness enhancement in the two-photon coincidence rate, supporting our hypothesis that it is an inherently better source, once properly compensated.

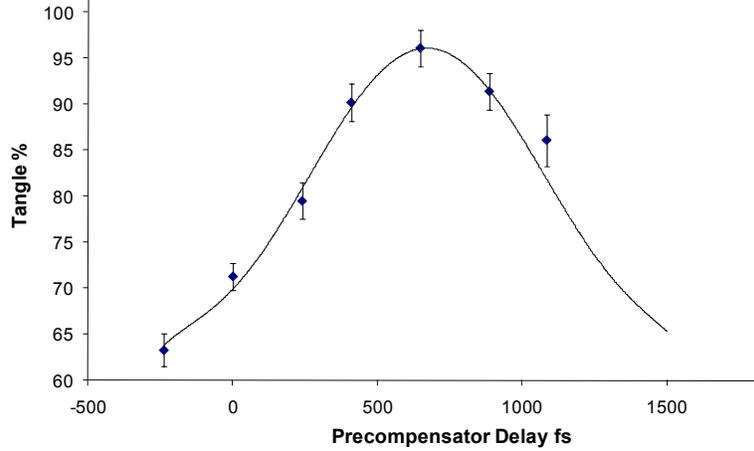

Fig. 3: Effect of varying the delay between the H- and V- pump components on the polarization-entangled state generated from two 0.6-mm BiBO crystals, pumped with a cw-diode laser for degenerate downconversion (405 nm $\rightarrow$ 810 nm + 810 nm). The delay is varied by using precompensators (in this case quartz) of differing lengths. A 0-fs precompensator delay represents the case when no temporal precompensator is present in the system. Negative delays, achieved by rotating the precompensator's orientation by 90°, add to the problem of temporal walkoff instead of countering it. The solid line is the theoretical prediction based on the theory presented in Section 3. The center of the peak is determined by the crystal lengths, index dispersion, and pump central wavelength; the peak width varies inversely with pump bandwidth.

## 3. Theoretical decoherence calculations and compensation design

While the theory for spatial and temporal distinguishability can be completely analyzed by fully expanding the downconversion state's dependence on $\omega$, $\theta$, and $\varphi$, in the spirit of [11, 37], here we present specific calculations necessary to design ultrafast and diode-laser pumped-entanglement sources. To make the following discussions as brief as possible, we outline the theoretical framework and refer to [21, 23, 24, 37, 38] for further details. To first order, spatial and spectral-temporal decoherence within the downconversion crystals themselves can be considered independent (our numerical simulations confirm this is a good approximation for the typical bandwidths and angles considered here) and hence, can be calculated separately (though as discussed above, the *compensators* for each will in general affect both).

### 3.1 Spatial decoherence and compensation

First, we calculate the emission-angle-dependent relative phase $\phi$ (see Eq. (1)) acquired by an entangled two-photon state generated by a monochromatic pump. Ordinary polarized [39] downconversion photons from the first crystal acquire an additional phase in the second crystal, where they are extraordinarily polarized, both because of spatial walkoff and the fact that they have to traverse the additional (i.e., the second) crystal. Three phase terms — the extraordinary phase $\Phi_e$, ordinary phase $\Phi_o$, and external phase $\Phi_\Delta$ — contribute to the total relative spatial phase in (1). Exact expressions for $\Phi_e$, $\Phi_o$, and $\Phi_\Delta$ can be found in [24]. To summarize, $\Phi_o$ accounts for the phase acquired by the ordinary downconversion photons in their birth crystal, and $\Phi_e$ arises because the downconversion photons born in the first crystal propagate through the second crystal as extraordinary photons. $\Phi_\Delta$ is the phase accumulated by a downconversion photon from the first crystal *outside* the second crystal (in which it is extraordinary polarized) relative to a photon created in the second crystal (ordinary polarized):

the former exits the second crystal at a different location than the latter due to spatial walkoff. The downconversion photons born in the first crystal acquire a net phase in the second crystal, given by

$$\phi_{dc}\left(\hat{\mathbf{k}}_{s,i}\right) = \left(\Phi_e + \Phi_\Delta\right)_{s,i} , \qquad (2)$$

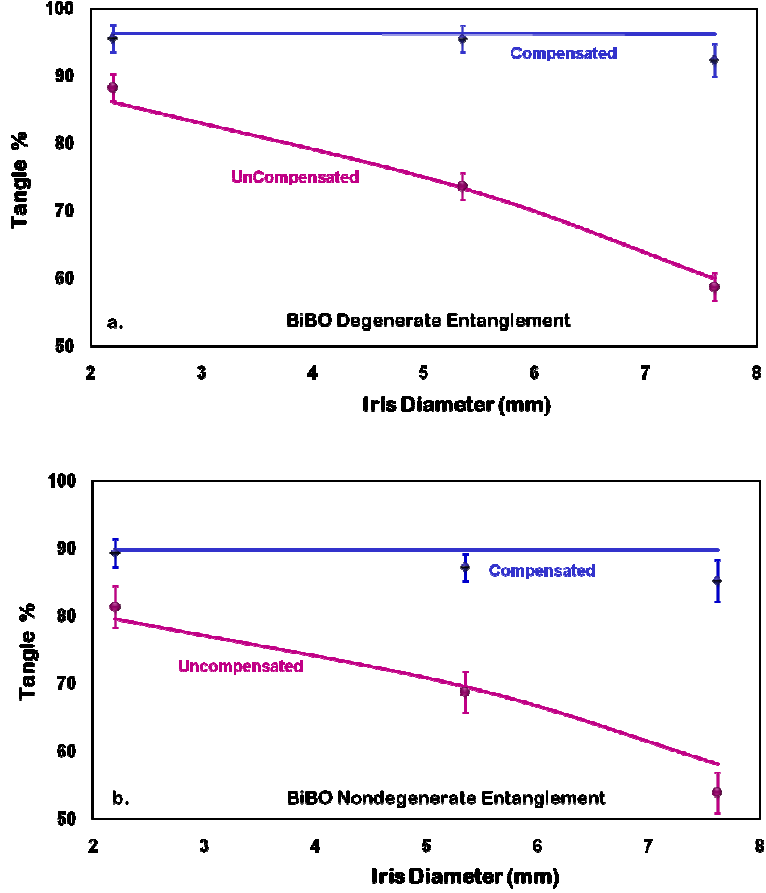

Fig. 4. Spatial compensation results for temporally compensated cw-laser diode-pumped BiBO entanglement source. **a**. degenerate downconversion (405 nm → 810 nm) and **b**. nondegenerate downconversion (405 nm → 851 nm + 772 nm). For this data, the collection irises were located ~84 cm from the downconversion source.

making the final polarization-entangled state

$$\left|\psi_{\hat{\mathbf{k}}_{s,i}}\right\rangle = \frac{1}{\sqrt{2}}\left(\left|H_1 H_2\right\rangle + e^{i\left(\phi_{dc}(\hat{\mathbf{k}}_s) + \phi_{dc}(\hat{\mathbf{k}}_i)\right)}\left|V_1 V_2\right\rangle\right). \qquad (3)$$

Multiple such states pass through a pair of finite-size collection irises; measuring these states together amounts to tracing over direction, producing the density matrix

$$\rho = \iint_{Iris} |\psi_{\hat{\mathbf{k}}_s}\rangle\langle\psi_{\hat{\mathbf{k}}_i}| d\hat{\mathbf{k}}_s d\hat{\mathbf{k}}_i. \qquad (4)$$

Averaging over the different relative phases leads to an effective decoherence, which can be compensated by appropriate birefringent crystals that impart an additional spatial phase opposite to that of the downconversion crystals. The polarization part of the final compensated state can be written in terms of the downconversion phases $\phi_{dc}$ (Eq. (2)) and the phases $\phi_c$ due to the spatial compensation crystals in the downconversion arms, as

$$\left|\psi_{\hat{\mathbf{k}}_{s,i}}\right\rangle = \frac{1}{\sqrt{2}}\left(|H_1 H_2\rangle + e^{i\left(\phi_{dc}(\hat{\mathbf{k}}_s)+\phi_{dc}(\hat{\mathbf{k}}_i)+\phi_c(\hat{\mathbf{k}}_s)+\phi_c(\hat{\mathbf{k}}_i)\right)}|V_1 V_2\rangle\right). \qquad (5)$$

When $\phi_{dc}(\hat{\mathbf{k}}_s) + \phi_{dc}(\hat{\mathbf{k}}_i) + \phi_c(\hat{\mathbf{k}}_s) + \phi_c(\hat{\mathbf{k}}_i) = 0$ (or any constant phase), the polarization part of the state factors out of the integral in (4), so that decoherence is suppressed [35]. Thus, spatial compensators allow increased brightness of the source, while maintaining high polarization-entanglement fidelity.

*3.2 Spectral-temporal decoherence and compensation: temporal domain analysis*

The second source of decoherence originates from the presence of different frequency components and can be analyzed in the spectral domain or equivalently, in the temporal domain (the two are related by a Fourier transform). For ease of calculations, we first present our compensation design analysis in the temporal domain here, and in Section 3.3 use a spectral domain picture to address more general considerations, e.g., the effect of spectral filtering. Spectral-temporal decoherence can be considered to arise because the emission times of the photons — specifically, *when* they exit the output face of the second crystal, relative to each other and/or relative to the time the pump photons entered the first crystal — depend on their frequencies and polarization because of dispersion and birefringence in the crystals. Assuming that the downconversion photons are emitted from the center of each crystal (see [38] for justification) the group-velocity dispersion effects which result in advancing or delaying the photons can be calculated as follows [23, 38]. The two-photon downconverted state from a single crystal can be written in the temporal domain in terms of creation operators acting on the vacuum state as

$$|\Psi\rangle = \int dt_s \int dt_i f(t_s,t_i) \hat{a}_s^+(t_s)\hat{a}_i^+(t_i)|vac\rangle, \quad (6)$$

where $f$, the joint two-photon amplitude (JTPA) which determines the signal-idler relationship, can be written for a pump with central frequency $\Omega_p$ and bandwidth $\sigma_p$ in terms of the spectral detunings $\nu_{s,i} \equiv \omega_{s,i} - \Omega_p/2$ and the phasematching function $\Theta$ as

$$f(t_s,t_i) = \frac{e^{-i\Omega_p \frac{t_s+t_i}{2}}}{2\pi} \iint d\nu_s d\nu_i e^{-i(\nu_s t_s + \nu_i t_i)} e^{-\left(\frac{\nu_s+\nu_i}{\sigma_p}\right)^2} \Theta(\nu_s,\nu_i). \quad (7)$$

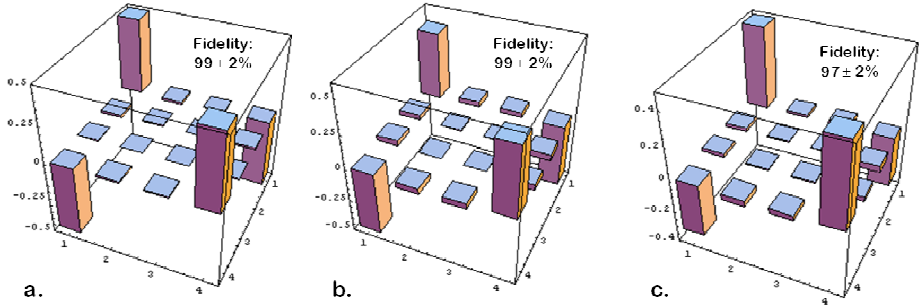

Fig.5. Density matrices of temporally and spatially compensated two-qubit polarization-entangled states produced by **a**. BBO pumped with 90-fs Ti-Saph, **b**. degenerate and **c.** nondegenerate downconversion in BiBO pumped by cw-diode laser (with 0.5-nm bandwidth).

**Table 1. Comparative two-photon coincidence brightness between BBO and BiBO crystals.**

| BBO (0.6mm) | BiBO (0.6 mm) |
|---|---|
| 5400 coincidences/mW/s | 13400 coincidences/mW/s |
| 0.75% background | 1.5% background |

The net time delay between the downconversion photons emitted in the first crystal $t_{s,i}^1$ (second crystal $t_{s,i}^2$) relative to when the pump is incident on the face of the first crystal can be written in terms of the group velocity $V_g^{ex}\left(V_g^{or}\right)$ of the extraordinary (ordinary) photons and crystal thickness $d$ as

and
$$t_{s,i}^1 = \frac{d}{2}\left(\frac{1}{V_g^{ex}(\omega_p)} + \frac{1}{V_g^{or}(\omega_{s,i})} + \frac{2}{V_g^{ex}(\omega_{s,i})}\right),$$

$$t_{s,i}^2 = \frac{d}{2}\left(\frac{2}{V_g^{or}(\omega_p)} + \frac{1}{V_g^{ex}(\omega_p)} + \frac{1}{V_g^{or}(\omega_{s,i})}\right). \quad (8)$$

For example, the first term in (8) represents the propagation of the pump photon through the first half of the first crystal, the second term represents the propagation of the signal or idler photon through the rest of the first crystal, and the final term represents the downconversion propagation through the second crystal. The net delays between the downconverted states from each crystal can be calculated for both the signal and idler as

$$\Delta t_{s,i}^{dc} \equiv t_{s,i}^2 - t_{s,i}^1 = d\left(\frac{1}{V_g^{or}(\omega_p)} - \frac{1}{V_g^{ex}(\omega_{s,i})}\right). \quad (9)$$

After both crystals the two-photon state in the time domain becomes

$$|\psi\rangle = \frac{1}{\sqrt{2}}\left(f(t_s+\Delta t_s^{dc}, t_i+\Delta t_i^{dc})|H,t_s\rangle_s |H,t_i\rangle_i + f(t_s,t_i)|V,t_s\rangle_s |V,t_i\rangle_i\right). \quad (10)$$

The downconversion polarization is now correlated with the time variable, leading to distinguishing which-crystal information. The effective density matrix associated with a polarization-entangled state can be obtained by tracing over the time variable and can be explicitly written as

$$\rho = \frac{1}{2}\left(|HH\rangle\langle HH| + |VV\rangle\langle VV| + v(\Delta t_s^{dc}, \Delta t_i^{dc})|HH\rangle\langle VV| + v^*(\Delta t_s^{dc}, \Delta t_i^{dc})|VV\rangle\langle HH|\right), \quad (11)$$

where $v(\Delta t_s^{dc}, \Delta t_i^{dc})$ is the convolution of the JTPA, defined as

$$v(\Delta t_s^{dc}, \Delta t_i^{dc}) \equiv \iint dt_s dt_i f\left(t_s + \Delta t_s^{dc}, t_i + \Delta t_i^{dc}\right) f^*\left(t_s, t_i\right). \tag{12}$$

Because $|HH\rangle$ and $|VV\rangle$ arrive at different times, the two-photon state is effectively decohered in the polarization basis. Mathematically, similar to spatial decoherence, such spectral-temporal decoherence can also be fully compensated for by introducing an appropriate birefringent element in the pump path. Ideally, the delay $\tau^{pc}$ introduced by a birefringent precompensator of length $l^{pc}$ depends on the respective group velocities $V_{g,pc}^{ex}$ and $V_{g,pc}^{or}$ of the extraordinary and ordinary polarizations in the precompensator:

$$\tau^{pc} = l^{pc}\left(\frac{1}{V_{g,pc}^{or}(\omega_p)} - \frac{1}{V_{g,pc}^{ex}(\omega_p)}\right). \tag{13}$$

The temporal precompensater shifts the extraordinary and ordinary pump components temporally so as to eliminate emission-time distinguishability between $|HH\rangle$ and $|VV\rangle$ i.e., $\tau^{pc} = -\Delta t_{s,i}^{dc}$. When $\Delta t_s^{dc}$ is significantly different from $\Delta t_i^{dc}$, which could be the case for downconversion at extremely nondegenerate wavelengths, then a post-downconversion crystal compensator is needed to completely compensate for the spectral-temporal decoherence [40]. Spatial compensators (length $l^{sc}$) placed in the path of the downconversion beams also contribute to the temporal delay, depending on the respective group velocities $V_{g,sc}^{ex}$ and $V_{g,sc}^{or}$ of the extraordinary and ordinary polarizations in the spatial compensators:

$$\tau_{s,i}^{sc} = l^{sc}\left(\frac{1}{V_{g,sc}^{or}(\omega_{s,i})} - \frac{1}{V_{g,sc}^{ex}(\omega_{s,i})}\right). \tag{14}$$

The final state can be written in terms of the precompensator delay $\tau^{pc}$ and spatial compensator temporal delay $\tau_{s,i}^{sc}$ by replacing $v(\Delta t_s^{dc}, \Delta t_i^{dc})$ in Eq. (11) with $v(\Delta t_s^{dc} - \tau^{pc} + \tau_s^{sc}, \Delta t_i^{dc} - \tau^{pc} + \tau_i^{sc})$. When $\tau^{pc} - \tau_{s,i}^{sc} = \Delta t_{s,i}^{dc}$, the temporal part of Eq. (10) factorizes from the polarization part, i.e., $|HH\rangle$ and $|VV\rangle$ become completely indistinguishable and one effectively recovers perfect polarization entanglement. Note that we chose to precompensate the (presumed collimated) pump as opposed to postcompensating (the non-collimated signal and idler photons) for the emission-time distinguishability, because precompensating a collimated beam results in no additional spatial dependence; in other words, while we need to account for the temporal effects of the spatial compensators, the spatial effects of the temporal precompensator are negligible in this case. Spatial and temporal calculations for biaxial crystals can be carried out similarly, using the slow and fast polarizations instead of the extraordinary and ordinary. In our numerical simulation, we based our biaxial phasematching calculations on the classical analysis by Beouf *et al.* [41].

*3.3. Spectral-temporal decoherence: spectral domain analysis*

The effects of emission-time distinguishability can be modeled and understood by comparing the difference to the pump coherence time as above; however, this simple model fails to intuitively explain why any effective decoherence occurs when a cw (but non-monochromatic) pump is used, since there is no pump timing to act as a relative which-crystal "clock" (although the temporal calculations themselves are still effectively valid for both cw-diode and ultrafast lasers). Thus, even though our compensation design calculations are performed in the temporal domain, a more fundamental description in the spectral domain elucidates the underlying physics, including why narrowband spectral filters can minimize these unwanted correlations. In the spectral domain the joint two-photon amplitude can be expressed as

$$f(v_s, v_i) = e^{-i\frac{d}{2}\left(D_+(v_s+v_i)+\frac{1}{4}D''(v_s-v_i)^2\right)} \text{sinc}\left(\frac{d}{2}\left(D_+(v_s+v_i)+\frac{1}{4}D''(v_s-v_i)^2\right)\right), \quad (15)$$

$$\text{where } D_+ = \frac{1}{V_g^{or}(\Omega_p/2)} - \frac{1}{V_g^{ex}(\Omega_p)}, \quad D'' = \left.\frac{d^2 \mathbf{k}_s}{d\omega^2}\right|_{\Omega_p/2}.$$

Here $D_+$ gives the group velocity mismatch between the downconversion photons and the pump, and $D''$ represents the group velocity dispersion [39]. Each frequency component in the pump and the downconversion sees a different effective length of the crystals, depending on polarization, leading to an effective frequency-dependent relative phase ϕ in (1). As with the emission-angle distinguishability discussed previously, averaging over the phases leads to an effective state decoherence. Since the extent to which frequency-polarization correlations matter depends on the pump bandwidth, such effects can become significant even for cw-diode lasers, which can have relatively large bandwidths (e.g., due to mode-hopping), requiring a precompensator to eliminate the spectral-temporal decoherence. The reason narrow bandwidth spectral filters can be used to recover the polarization entanglement — at the cost of collection efficiency — is that the measured JTPA is given by the convolution of the JTPA before the filters, shown in (15), with the frequency response of the spectral filters. Spectral filters limit the collected bandwidth or equivalently, reduce the integration interval, and thus the amount of temporal decoherence. Extreme cases, e.g., ultrashort pulsed lasers (which necessarily have a very broad spectral bandwidth) require prohibitively narrow spectral filters to maintain high fidelity, making the source extremely inefficient without temporal precompensators.

### 4. Conclusion

Ultrabright sources of polarization-entangled photons are a critical resource for scalable optical quantum information processing. Type-I sources, especially those pumped by ultrafast-pulsed and cw-diode lasers, can be optimized by compensating for phase-decoherence mechanisms. Furthermore, the nonlinear downconversion crystals themselves can be upgraded, e.g., the brightness of polarization-entangled photons can be dramatically improved (a three-fold brightness improvement compared to BBO) by using BiBO, a highly nonlinear crystal. Here we have modeled and demonstrated for the first time high-fidelity (>99%) type-I polarization-entangled photons directly from BiBO. Further, we have combined for the first time two compensation techniques — spatial and spectral-temporal compensation — to counter the two limiting dephasing mechanisms, and report the highest fidelity (99%) maximally polarization-entangled state achieved using an ultrafast Ti-Saph laser. Our polarization-entanglement simulation code (freely available on our website: http://research.physics.illinois.edu/QI/photonics/phase_compensation.html), experimentally tested and reported here, can predict state quality and design compensation crystals for a variety of type-I sources. We hope that our compensation design code will benefit other researchers interested in employing such optimized entanglement sources.

### Acknowledgements

We acknowledge helpful contributions to our entangled source simulation from Gleb Akselrod, Joseph Altepeter, Jaime Valle and Joseph Yasi. This work was supported by the National Science Foundation (Grant # EIA-0121568), MURI Center for Photonic Quantum Information Systems (ARO/ARDA Program DAAD19-03-1-0199), and the IARPA-funded